\documentclass[preprint,12pt]{elsarticle}

\usepackage{amssymb}
\usepackage{amsmath}

\usepackage{gensymb}

\biboptions{sort&compress}

\usepackage{float}

\usepackage[T1]{fontenc}

\usepackage[utf8]{inputenc}

\usepackage{array} 

\usepackage{lineno}

\journal{Optics and Laser Technology}

\begin{document}

\begin{frontmatter}

\title{Design of Dual-Band Plasmonic Absorber for Biomedical Sensing and Environmental Monitoring}

\author[1,2]{Ayon Sarker\corref{cor1}} 
\cortext[cor1]{https://orcid.org/0000-0001-5808-4970}
\affiliation[1]{organization={Department of Electrical and Electronic Engineering, Bangladesh University of Engineering and Technology},
            city={Dhaka},
            country={Bangladesh}}

\affiliation[2]{organization={Department of Computer Science and  Engineering, BRAC University},
            city={Dhaka},
            country={Bangladesh}}

\author[1]{Sajid Muhaimin Choudhury\corref{cor2}\fnref{fn1}}
\cortext[cor2]{https://orcid.org/0000-0002-0216-7125}
\fntext[fn1]{sajid@eee.buet.ac.bd}

\begin{abstract}

\noindent This study introduces a dual-band plasmonic absorber designed for simultaneous sensing applications in the near-infrared (NIR) and mid-infrared (MIR) regions. The absorber, composed of silver nanostructures on a metal plate with a dielectric spacer, exhibits a combination of localized and gap surface plasmon resonances, resulting in two distinct absorption peaks in theoretical analysis based on the FDTD method. Numerical simulations also validate the sensor’s high refractive index sensitivity, enabling the detection of biomolecules, proteins, viruses, and various solutes in aqueous solutions. The absorber demonstrates significant resonance shifts, making it a promising candidate for environmental monitoring, medical diagnostics, and chemical sensing.

\end{abstract}

\begin{keyword}
Dual-band, Plasmonic absorber, RI sensing, Bio-chemical sensing, Environmental monitoring
\end{keyword}

\end{frontmatter}


\section{Introduction}
\label{sec:Introduction}
The interest in plasmonics, an intriguing field that focuses on the interactions between free electron clouds in metallic structures and incoming photons\cite{stockman2018roadmap}, has surged significantly, leading to remarkable progress, especially in sensor technology\cite{duan2021surface,lee2021quantum}. Plasmonic absorbers have emerged as powerful tools in nanophotonics due to their extraordinary ability to confine electromagnetic (EM) waves at subwavelength scales, thereby facilitating enhanced light-matter interactions\cite{divya2022surface,butt2024insight,ghobadi2022strong}. This capability has positioned them as a highly promising option for a range of applications, including bio-sensing, chemical sensing, gas sensing, biological analyte sensing, environmental and healthcare monitoring and sensing,  photodetection, energy harvesting, and thermal emission\cite{duan2021surface,li2023silver}. Additionally, the combination of high sensing efficiency, cost-effectiveness, compact device size, stability, and the remarkable variety, accessibility, and reusability of optically functional materials has sparked significant interest in plasmon-based sensing technologies\cite{sudarsan2012optical}.

The driving principle behind plasmonic absorbers is surface plasmon resonance (SPR)\cite{jang2016plasmonic}, which refers to the excitation of surface plasmons that occurs when incident EM waves induce collective oscillations of free electrons at the metal-dielectric interface\cite{10.1039/9781847558220-00015}. SPRs involve two fundamental types of electron oscillations triggered by incident EM radiation at metal-dielectric interfaces: surface plasmon polaritons (SPPs) and localized surface plasmons (LSPs)\cite{du2010localized}. SPPs propagate along a conductor's surface, while LSPs, confined to metallic nanostructures, do not propagate but are closely coupled with the EM field. While LSPs can be excited directly by incidenting light, SPPs, on the other hand, require specific phase-matching techniques\cite{tang2019concept, zalyubovskiy2012theoretical}. Due to their exceptional sensitivity to changes in the surrounding refractive index, refractive index sensors based on both propagating surface plasmon resonance (PSPR)\cite{rao2024self, hu2023universal} and localized surface plasmon resonance (LSPR)\cite{dormeny2020design, singh2020numerical} have been developed, each offering its advantages and limitations\cite{jatschka2016propagating}.

While LSPR sensors offer tunability based on nanoparticle size and shape\cite{min2020manipulating}, their sensitivity is generally lower than PSPR sensors\cite{wang2023sensitivity}. However, to enhance sensor performance, a combination of these two modes can be utilized, particularly when using periodic nanoparticle arrays\cite{wang2023sensitivity, dang2020efficient}. This is due to the energy transfer resulting from the enhanced electromagnetic coupling between LSP and PSP\cite{liu2012controllable}, which takes place when a metallic nanoparticle is in the close vicinity of a metallic surface through the interaction of dipoles within it and their mirror image on the metallic surface\cite{farhang2013coupling}. Moreover, the strength of interparticle coupling can be enhanced to a great extent by employing a periodic or quasi-periodic arrangement of nanoparticles (metasurfaces) instead of a single nanoparticle\cite{bukhari2019metasurfaces}. Another type of surface plasmon confinement, known as gap plasmon, occurs in the gap between two surfaces in a metal-insulator-metal (MIM) configuration—a 2D planar array supported by a metallic film with a thin spacer\cite{chou2021significantly}. Combining these phenomena within a single setup provides superior control over reflected waves, enhances spectral selectivity, and enables nonlinear optical properties, all while maintaining a compact and streamlined device footprint\cite{chou2020perfect,liang2022ultra}.

Plasmonic absorbers are greatly admired, and researchers have been harnessing their advantages in the infrared (IR) region of the electromagnetic spectrum\cite{zhong2015review}. This is because IR light has a unique ability to resonate with most molecules due to the transition of molecular vibration and thus exhibits exceptional diffraction, penetration, and absorption properties\cite{sandorfy2007principles}, making this band essential for sensing and spectroscopy applications in medical, biotechnology, security, and defense fields\cite{mayerhofer2021recent}. The span of the IR spectrum is conventionally subdivided into three regions: near-infrared (NIR) (780–2500 nm); mid-infrared (MIR) (2500–25,000 nm); and far-infrared (FIR), which includes wavelengths beyond 25,000 nm\cite{kontsek2020mid}. NIR wavelengths are effective for detecting organic molecules, proteins, and proteinaceous structures\cite{manley2014near}, while MIR wavelengths are utilized mostly to identify specific molecular signatures\cite{de2018applications}. Despite the numerous distinct conveniences that can be achieved in NIR and MIR bands individually, most reported IR absorbers have been limited to functioning within just one of these spectral ranges, which narrows their application scope and diminishes their potential effectiveness for a wide variety of sensing and detection tasks. Introducing the concept of developing dual-band plasmonic absorbers can offer a compelling solution to this limitation. By operating across different spectral regions, this type of plasmonic absorber can enhance its functionality and be capable of simultaneous detection of multiple analytes or even various properties of a single analyte. Additionally, they can be particularly valuable in applications requiring high sensitivity across diverse wavelengths, such as environmental monitoring, medical diagnostics, and chemical sensing. Nevertheless, despite recent advancements in designing dual-band IR plasmonic absorbers for sensing, filtration, and characterization applications, most of their resonant frequencies still fall exclusively within either the NIR\cite{alipour2020ultra, chou2021biosensing, cheng2018dual} or MIR\cite{chen2012dual, zhang2023dual, chen2024dual} range, leaving the aforementioned concern unresolved. 

Taking this issue into account, this study introduces a dual-band narrowband absorber consisting of arrays of silver nanostructures on a metal plate, with a layer of polymer separating them. The structure combines the effects of SPPs and LSPs, resulting in two absorption peaks: one at 1366 nm (219.5 THz, 7320.64 $ \text{cm}^{-1} $) in the NIR region—specifically in the NIR-III range (1350 nm–1870 nm\cite{shen2022defect}; allowing for deeper penetration in biological tissues/fluids with minimal scattering, making this range more suitable for biological and biochemical detection in particular\cite{shen2022defect,sordillo2014deep}) with an absorption rate of 99.3\% and another in the MIR region with 99.9\% absorption at 2683 nm (111.74 THz, 3727.17 $ \text{cm}^{-1} $). The simulation data from the numerical analysis of the proposed structure, using the finite-difference time-domain (FDTD) method, validate its ability to achieve dual resonances simultaneously and confirm its suitability as a plasmonic sensing platform, with the resonance being sensitive to the refractive index of the surrounding media. This sensitivity has been thoroughly examined through simulations for various applications, including the detection of gas hydrates in seawater, determining concentrations of solvents or the presence of biomolecules in aqueous solutions and organic fluids, detection of immobilized bio-layers, confirmation of DNA hybridization completion, etc. Due to its relatively simple structure and the compelling ability to operate simultaneously in two distinct regions of the IR spectrum, this absorber can be an attractive candidate for next-generation sensing technologies in medical diagnostics, environmental science, and chemical detection.

\section{Structural Composition and Layout}
\label{sec:Structural Composition and Layout}
\begin{figure}[H]
    \centering
    \includegraphics[width=1\textwidth]{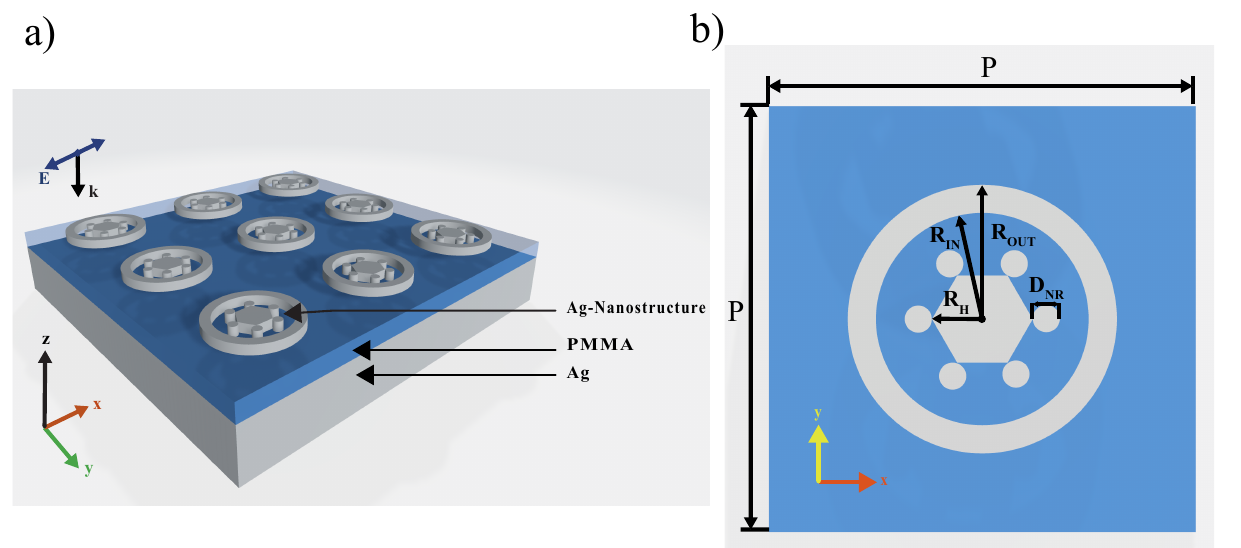}
    \caption{a) Three-dimensional illustration of the proposed absorber, b) Orthographic top view of the unit cell.}
    \label{fig:Fig-1}
\end{figure}

The structure consists of three layers, starting with a metal plate at the base. A two-dimensional square array of metal nanostructure is deposited as the top layer on a sheet made of dielectric material that acts as a spacer between the top and bottom metal layers. Figure \ref{fig:Fig-1}.a) illustrates the proposed layout of our plasmonic absorber. We used silver (Ag) to prepare the top layer of nanostructures and the metal plate at the bottom since it has significantly lower interband optical loss compared to other commonly used plasmonic metals like gold, aluminum, and copper\cite{boltasseva2011low}. The thickness of the Ag-nanostructures is 45 nm, and that of the silver plate at the base is 265 nm. We selected the thickness of the bottom plate to prevent any transmission of the incident light, taking into consideration the frequency-dependent skin depth of silver within the spectrum we are operating in\cite{wu2021four}. A 90 nm thick layer of polymethyl methacrylate (PMMA) is incorporated on top of the silver layer as the dielectric spacer in this absorber. On top of this polymer layer, the square array of Ag-nanostructures is periodically repeated in the x and y directions. The Ag-nanostructures can be fabricated on the PMMA layer using the techniques reported previously\cite{singh2007situ, kazemian2011electrical}. 

\noindent Figure \ref{fig:Fig-1}.b) shows a unit cell with a side length of P = 675 nm with a detailed view of the Ag-nanostructure. Each Ag-nanostructure contains a ring made of silver with inner radius $R_{IN}$ = 170 nm and outer radius $R_{OUT}$ = 210 nm. At the center of the ring, there is a hexagonal cuboid with a circumradius of $R_H$ = 80 nm. Lastly, we placed six Ag-nanorods with a diameter of 40 nm at the vertices of the hexagon. The locations of the centers of the nanorods arranged around the central hexagon are calculated such that the walls of the nanorods just make contact with the vertices of the hexagon. This is achieved by utilizing polar coordinates and applying the following expressions \ref{equation:equation-1}, and \ref{equation:equation-2},

\begin{equation}
    r_n = R_H + \frac{D_{NR}}{2}
    \label{equation:equation-1}
\end{equation}

\begin{equation}
    \theta_n = \frac{\pi}{3} \cdot n  
    \label{equation:equation-2}
\end{equation}

\noindent Where n = 0,1,2,3,4, and 5. 

The square unit cell shown in Figure \ref{fig:Fig-1}.b) containing one Ag-nanostructure each on top of the PMMA layer, and a silver plate underneath, is two-dimensionally repeated to fabricate the entire layout of the proposed structure. 

\section{Methods}
\label{sec:Methods}
We simulated the proposed structure numerically by applying the three-dimensional finite difference time domain (FDTD) method. To model our structure for the simulation we employed the refractive index data of silver from\cite{palik1998handbook}, and for PMMA, using the Sellmeier model, we derived and utilized values calculated from the equation \ref{equation:equation-3} below\cite{paschotta2008encyclopedia,polyanskiy2024refractiveindex},

\begin{equation}
    n^2-1 = \frac{0.99654\lambda^2}{\lambda^2-0.00787} + \frac{0.18964\lambda^2}{\lambda^2-0.02191} + \frac{0.00411\lambda^2}{\lambda^2-3.85727}
    \label{equation:equation-3}
\end{equation}

To obtain the absorption spectrum, we conducted simulations to calculate the reflection and transmission spectra by illuminating the absorber with a plane wave that is normally incident, propagating along the negative z-axis, and applying suitable boundary conditions along all three axes. In simulations periodic boundary conditions are imposed on all four sides of the structure in the x and y directions, while a perfectly matched layer (PML) absorption condition is applied to the top and bottom in the direction of the incident illumination. By analyzing absorption spectra computed with proper simulation settings for different applications, we also investigated the figure of merit (FOM), refractive index sensitivity, the full width at half maximum (FWHM), and other properties of the absorber we introduced in this work.

\section{Result and discussion}
\label{sec:Result and discussion}
The introduced absorber structure exhibits two absorption peaks when illuminated by light from a plane source. Utilizing the data from simulations, the absorption spectrum can be assessed using the expression A = 1-R-T, where R and T refer to reflection and transmission, respectively. Since the thickness of the silver plate at the base is chosen to be greater than its skin depth, the transmission is virtually zero all over the spectrum. 

\begin{figure}[H]
    \centering
    \includegraphics[width=1\textwidth]{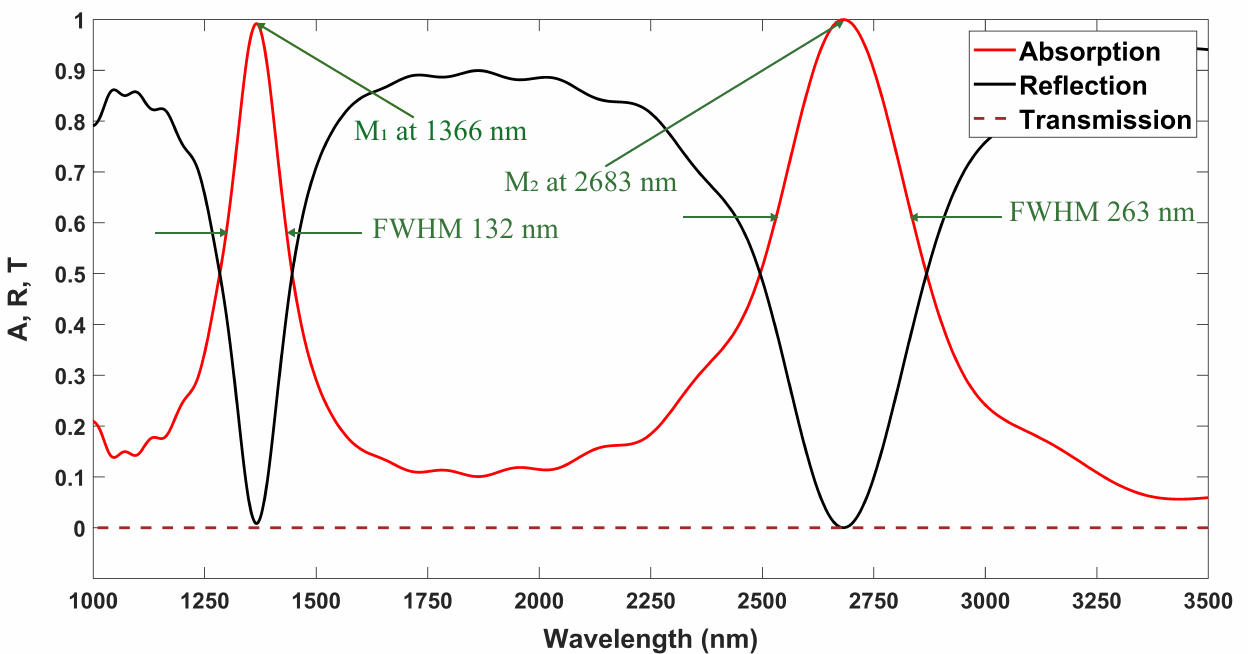}
    \caption{Reflection, absorption, and transmission spectra of the proposed absorber for normally incident TM-polarized light with the electric field along the x-axis. The low transmission indicates high absorption at the resonant frequency, as A = 1 - R - T.}
    \label{fig:Fig-2}
\end{figure}

\noindent From Figure \ref{fig:Fig-2}, it can be noticed that there are two dips in the reflection spectrum implying two absorption peaks—one at 1366 nm in the NIR-III range (Mode -1 or $M_1$) with 99.5\% absorption (reflection 0.46\%) and the second one at 2683 nm with 99.99\% absorption in the MIR band (Mode-2 or $M_2$) (reflection 0.0051\%). 

\begin{figure}[H]
    \centering
    \includegraphics[width=1\textwidth]{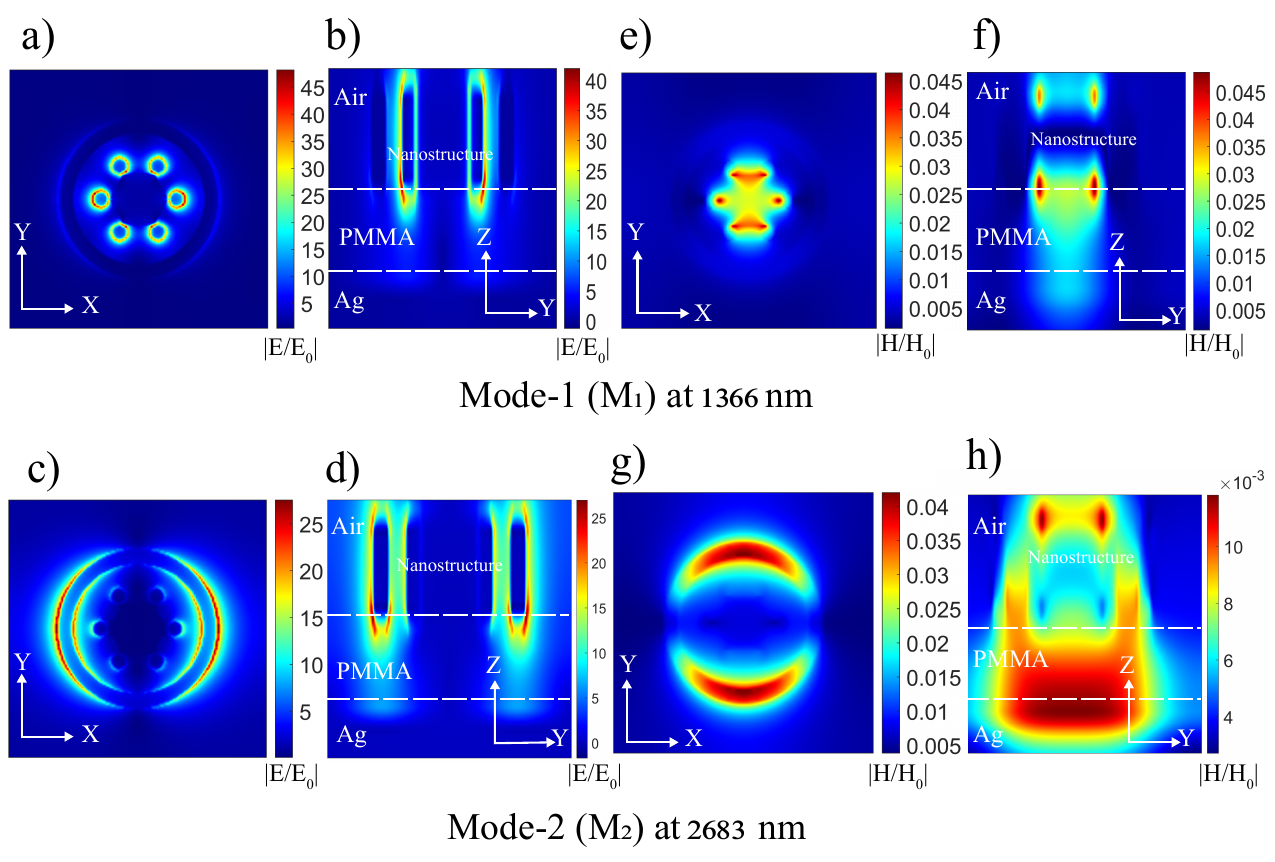}
    \caption{Electric and magnetic field distribution in the structure, a) Localized electric field distribution at the XY cross-section at the Ag-nanostructure and spacer interface for $M_1$, b) Localized electric field distribution at the YZ cross-section for $M_1$, c) Localized electric field distribution at the XY cross-section at the Ag-nanostructure and spacer interface for $M_2$, d) Localized electric field distribution at the YZ cross-section for $M_2$, e) induced magnetic field distribution at the XY cross-section at the Ag-nanostructure and spacer interface for $M_1$, f) induced magnetic field distribution at the YZ cross-section for $M_1$, g) induced magnetic field distribution at the XY cross-section at the Ag-nanostructure and spacer interface for $M_2$, h) induced magnetic field distribution at the YZ cross-section for $M_2$.}
    \label{fig:Fig-3}
\end{figure}

To investigate the underlying causes of the absorber's behavior at resonance frequencies, we thoroughly examined the electromagnetic field induced by the incident light within the absorber at two resonance wavelengths separately as shown in Figure \ref{fig:Fig-3}.

\noindent For $M_1$, Figure \ref{fig:Fig-3}.a) and b) show the localized electric field that forms LSPR around the nanorods of the nanostructures on the topmost layer. The LSPR strongly couples to its mirror dipole on the inner wall of the silver ring and to the one on the bottom silver plate. In the case of $M_2$, as seen in Figure \ref{fig:Fig-3}.c), and d) the electric field localizes at the outer and inner walls of the ring, and similarly couples to its dipole even more strongly than that for $M_1$.  At both resonant frequencies, gap surface plasmon (GSP) can be noticed in the spacer layer of PMMA beneath the nanorods and ring\cite{ding2018review}.
Furthermore, from Figure \ref{fig:Fig-3}.e), and g)it can be observed that the induced magnetic field is concentrated at the hexagonal cuboid mainly in the Y direction for $M_1$, while for $M_2$, it is concentrated at the inner wall of the nanoring in the Y direction. Additionally, Figure \ref{fig:Fig-3}.f) and h) show the intensity of the confined magnetic field along the Z axis within the absorber for $M_1$ and $M_2$ respectively. The magnetic field induced by the incident light aids in the enhancement of the confined electromagnetic field resulting in low reflectivity and thereby greater absorption\cite{halas2011plasmons}. 

\begin{figure}[H]
    \centering
    \includegraphics[width=0.85\textwidth]{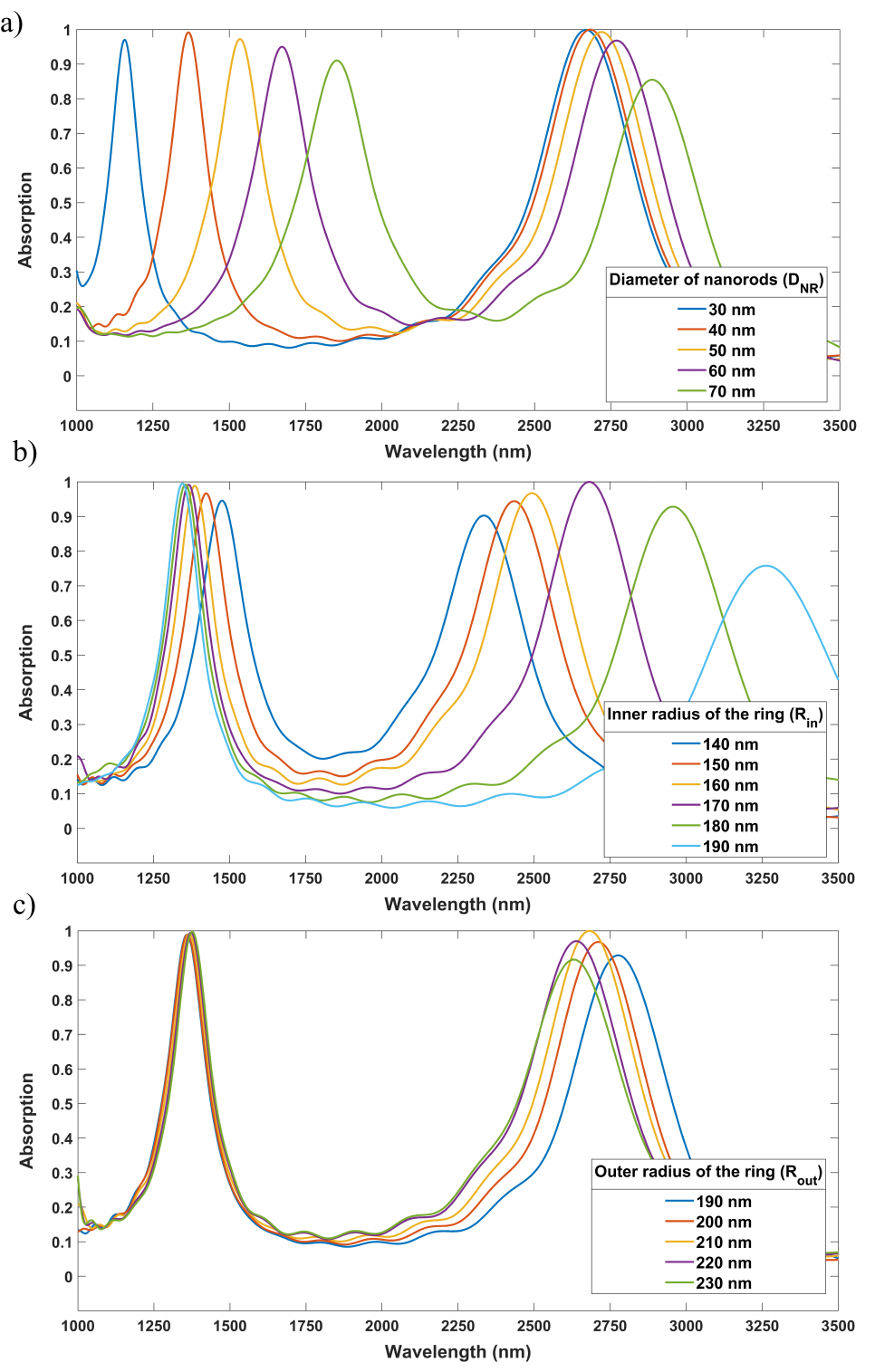}
    \caption{Change in absorption spectra in response to the change in a) diameter of nanorods ($D_{NR}$), b) Inner radius of the ring ($R_{IN}$), c) Outer radius of the ring ($R_{OUT}$).}
    \label{fig:Fig-4}
\end{figure}

To examine the impact of geometric parameters of the introduced Ag-nanostructure, first, we varied the diameter of the nanorods ($D_{NR}$).  With the increasing diameter of nanorods ($D_{NR}$), a redshift in both resonance wavelengths can be noticed in Figure a). The change in diameter of nanorods ($D_{NR}$) affects the position of the $M_1$ more since the resonance at $M_1$ is essentially associated with the nanorods as noticed from Figure \ref{fig:Fig-4}.a). Figure \ref{fig:Fig-4}.b), and c) show the consequence of the change in the inner and outer radii ($R_{IN}$ and $R_{OUT}$) of the nanoring. Since it is seen in Figure \ref{fig:Fig-3}.c) that the resonance at $M_2$ is linked to both the inner and outer walls of the ring, changes in the radii primarily influence the position of the $M_2$ resonance. However, as the inner radius of the ring ($R_{IN}$) increases, the LSPR between the inner wall and the nanorods couples less strongly due to the expanding gap between them, resulting in lower absorption and shifts in the positions of both $M_1$ and $M_2$. Alternatively, adjusting the outer radius of the ring ($R_{OUT}$) has minimal to no effect on the position of $M_1$. However, as the width of the ring varies, both the absorption rate and the position of $M_2$ change, as shown in Figure \ref{fig:Fig-4}.c).

With the altering values of the polarization angle of the incident light, the localized electric field profile in the structure remains constant, thereby the locations of the resonances and the amount of absorption stayed almost unchanged. This consistency is attributed to the highly symmetrical configuration of the proposed structure. 

\begin{figure}[H]
    \centering
    \includegraphics[width=0.9\textwidth]{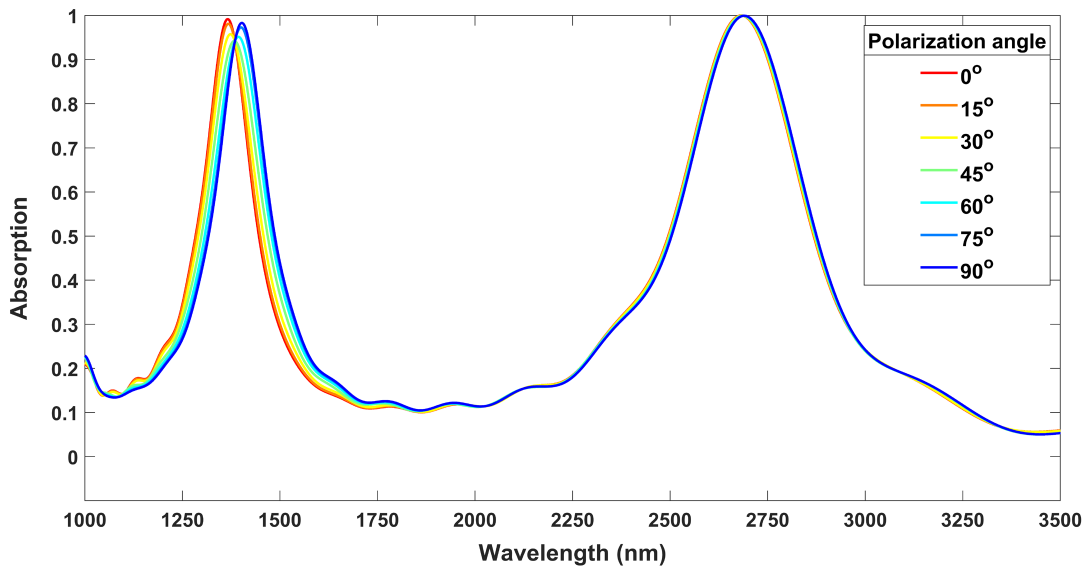} 
    \caption{Absorption spectra obtained for various polarization angles of the normally incident light.}
    \label{fig:Fig-5}
\end{figure}

\noindent Figure \ref{fig:Fig-5} shows the absorption spectra of the absorber when illuminated by incident light with different polarization angles individually. As the polarization angle varies from 0° to 90°, the position of $M_1$ shifts from 1366 nm to 1400 nm, while $M_2$ shows almost no change as it is associated with the ring of the nanostructure, which has complete radial symmetry and remains unaffected irrespective of the polarization angle.

The dependency of the absorption spectra of the structure for inclined indent of the light other than normal incident is also investigated. A detailed discussion and results are provided in the Supplementary article.

\subsection{Sensing performance}
\label{subsec:Sensing performance}
To assess the sensing performance of our introduced absorber as a refractive index sensor, first, we calculated the refractive index sensitivity ($RIS$). Refractive index sensitivity is a key and widely utilized parameter for assessing a sensor's performance. It measures how the sensor's resonance wavelength responds to changes in the bulk refractive index of the environment in which it operates and can be expressed through the equation below\cite{xu2019optical},

\begin{equation}
    RIS = \frac{\Delta \lambda_{\text{res}}}{\Delta n}.
\end{equation}
Here, \(\Delta n\) represents the change in refractive index, while \(\Delta \lambda_{\text{res}}\) denotes the corresponding change in resonance wavelength.\\

\begin{figure}[H]
    \centering
    \includegraphics[width=1\textwidth]{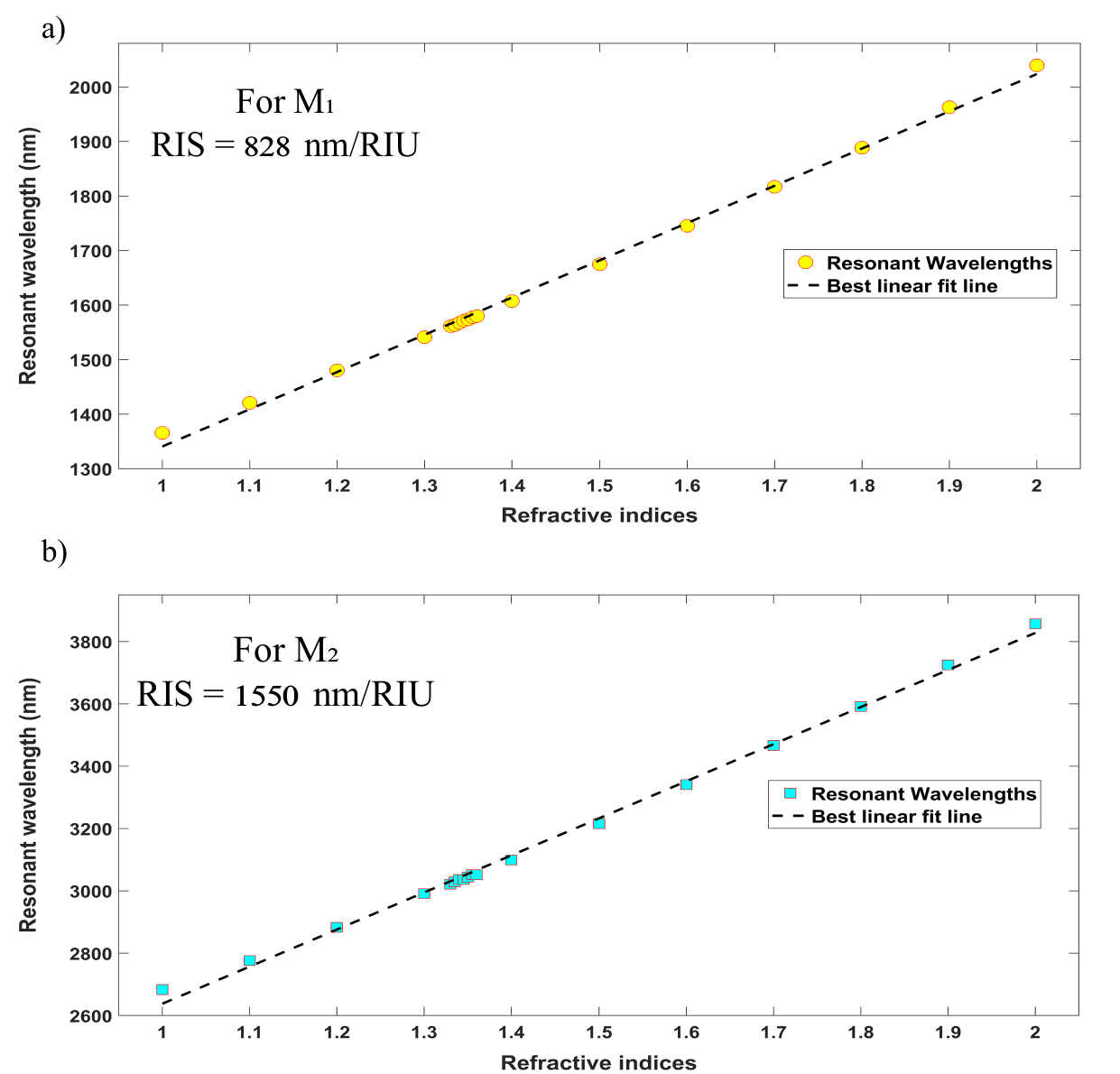}
    \caption{The resonant wavelengths of the two modes exhibit a linear trend in shifting as they change with the effective refractive index of the surrounding environment a) for $M_1$, b) for $M_2$}
    \label{fig:Fig-6}
\end{figure}

\noindent We adjusted the refractive index of the sensor's surrounding environment from 1.0 to 2.0 in increments of 0.1 and determined the resonance wavelengths for both modes individually for each instance as shown in Figure \ref{fig:Fig-6}.a), and b). Additionally, we considered the values of the refractive index from 1.33 to 1.36 with a step size of 0.005, since analytes in biosensing applications typically have a refractive index within this range\cite{wang2017label,li2020review}. Analyzing the data obtained from these simulations, we found the RIS value for M1 to be 828 nm/RIU, and that for M2 to be 1550 nm/RIU, which is satisfactorily high for biosensing and chemical sensing purposes. A linear redshifting trend of the resonance wavelengths can also be noticed in Figure \ref{fig:Fig-6}.a), and b).

We also computed the figure of merit of our absorber in sensing tasks considering both $M_1$ and $M_2$. The figure of merit ($FoM$) standardizes the bulk refractive index sensitivity ($RIS$) by relating it to the resonance curve's width, defined by the full width at half maximum ($FWHM$), which determines the accuracy of measuring the resonance minimum\cite{xu2019optical}.

\begin{equation}
    {FoM} = \frac{RIS}{FWHM}.
\end{equation}

\noindent The equation above that describes FoM was utilized to calculate the values of FoM, and the obtained values are 6.3 per RIU and 5.9 per RIU for $M_1$ and $M_2$, respectively.



\subsection{Comparison with literature}

\begin{table}[h!]
\centering
\renewcommand{\arraystretch}{1.5} 
\begin{tabular}{>{\centering\arraybackslash}m{3cm}>{\centering\arraybackslash}m{4cm}>{\centering\arraybackslash}m{3cm}}
\hline
\hline
Reference & Operating band & $RIS_{max}$ \\ 
\hline
\hline
\cite{zahra2024design}  & Only near-infrared & 733 nm/RIU     \\ 
\cite{amoosoltani2021double}  & Only near-infrared & 1000 nm/RIU    \\ 
\cite{madadi2020dual}  & Only near-infrared & 652.77 nm/RIU  \\ 
\cite{chou2021biosensing}   & Only near-infrared & 1000 nm/RIU   \\ 
\cite{alipour2020ultra}  & Only near-infrared & 1240.8 nm/RIU \\ 
\cite{chen2021multi}  & Only Mid-infrared & 907.88 nm/RIU \\ 
\cite{chen2022multi}  & Only Mid-infrared & 839.39 nm/RIU \\ 
This work & Both near-infrared and mid-infrared & 1550 nm/RIU   \\ 
\hline
\hline
\end{tabular}
\caption{Comparison with dual/multi-band sensors  reported in recent years}
\label{table:Tab-1}
\end{table}

We compared the evaluated sensing performance with previously reported dual or multi-band absorbers, as shown in Table \ref{table:Tab-1}. It is evident that the absorber introduced in this study offers moderately higher refractive index sensitivity, while simultaneously operating in two distinct ranges of the IR spectrum, thereby broadening its potential range of applications.



\section{Applications}
\label{sec: Applications}

To examine the compatibility of our proposed absorber as a sensor in the field of biosensing and environment monitoring, we incorporated it into diverse sensing scenarios. By applying the principle of refractive index sensing, we simulated its use for detecting gas hydrates, identifying various proteins at different concentrations dissolved in water, distinguishing amino acids from their isomers, and determining concentrations of various inorganic and organic solutes in aqueous solutions to assess its performance as a chemical sensor. Additionally, to investigate its potential as a biosensor, we simulated its application in detecting E. coli in food and beverage samples, quantifying biomolecule attachment on its surface, measuring hemoglobin concentration in human blood, confirming DNA hybridization completion, and detecting various viruses. Detailed results and discussions are provided in the Supplementary article.



\section{Conclusion}

To summarize, the dual-band plasmonic absorber we presented in this letter exhibits high sensitivity across both the NIR and MIR regions, offering the potential for various sensing applications such as biomolecule detection, virus identification, and environmental monitoring. The structure’s ability to achieve high refractive index sensitivity, along with its compact design and compatibility with a broad range of analytes, implies its promise for advanced sensing technologies. Simulations demonstrated that the absorber’s performance is enhanced by combining localized and gap surface plasmon resonances, leading to significant shifts in resonance wavelengths with changes in the surrounding environment. This makes it a versatile platform for a wide range of practical applications.










\bibliographystyle{elsarticle-num}  
\bibliography{References}  

\end{document}


\begin{frontmatter}
\title{Design of Dual-Band Plasmonic Absorber for Biomedical Sensing and Environmental Monitoring: supplementary material}
\end{frontmatter}

\section{Result and discussion}
\label{sec: Result and discussion}
\subsection{Varying the angle of incident of the light}

\begin{figure}[H]
    \centering
    \includegraphics[width=1\textwidth]{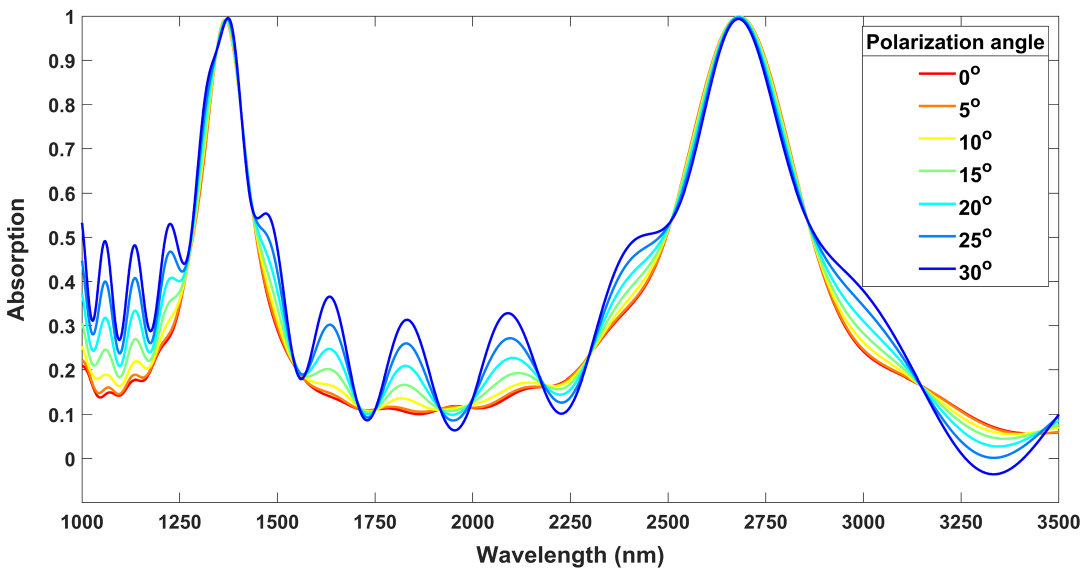}
    \captionsetup{labelformat=empty, labelsep=none}
    \caption{Figure S1: Absorption spectra obtained for various incident angles of the incident light.}
    \label{fig:Fig-s1}
\end{figure}

Figure S\ref{fig:Fig-s1} shows the comparison of the structure's behavior under light at various incident angles rather than at normal incidence. As illustrated, the resonance wavelengths for both modes remain unchanged regardless of the incident angle, as does the absorption rate. However, the reflectance fluctuates slightly at non-resonance frequencies for larger incident angles, which can be conveniently resolved by illuminating the absorber with light at angles less than 20$\degree$.


\section{Applications}
\label{sec: Applications}

\subsection{Detection of gas hydrate}

The proposed structure was simulated to determine its potential application in assessing the presence of gas hydrates in seawater using the concept of refractive index sensing. Clathrate hydrates, commonly known as gas hydrates, are crystalline compounds formed from water and gases under thermodynamically favorable conditions\cite{englezos1993clathrate}. Determination of the existence of gas hydrates in seabed water is crucial for gas and fuel industries as it implies the potential existence of gas fields beneath the sea bed \cite{andersson2005gas}. Apart from this, since gas hydrates are being explored in various applications such as storing gas safely\cite{veluswamy2021natural}, and cold energy\cite{sun2016review}, the step of determining the presence of gas hydrates has become important. To detect the existence of gas hydrate, we simulated the structure for two different values of ambient refractive index and measured the shift in resonant wavelengths. The refractive index of water and gas hydrate in water at $1\degree$ C are 1.33 and 1.346, respectively\cite{bylov1997experimental}. 

\begin{figure}[H]
    \centering
    \includegraphics[width=1\textwidth]{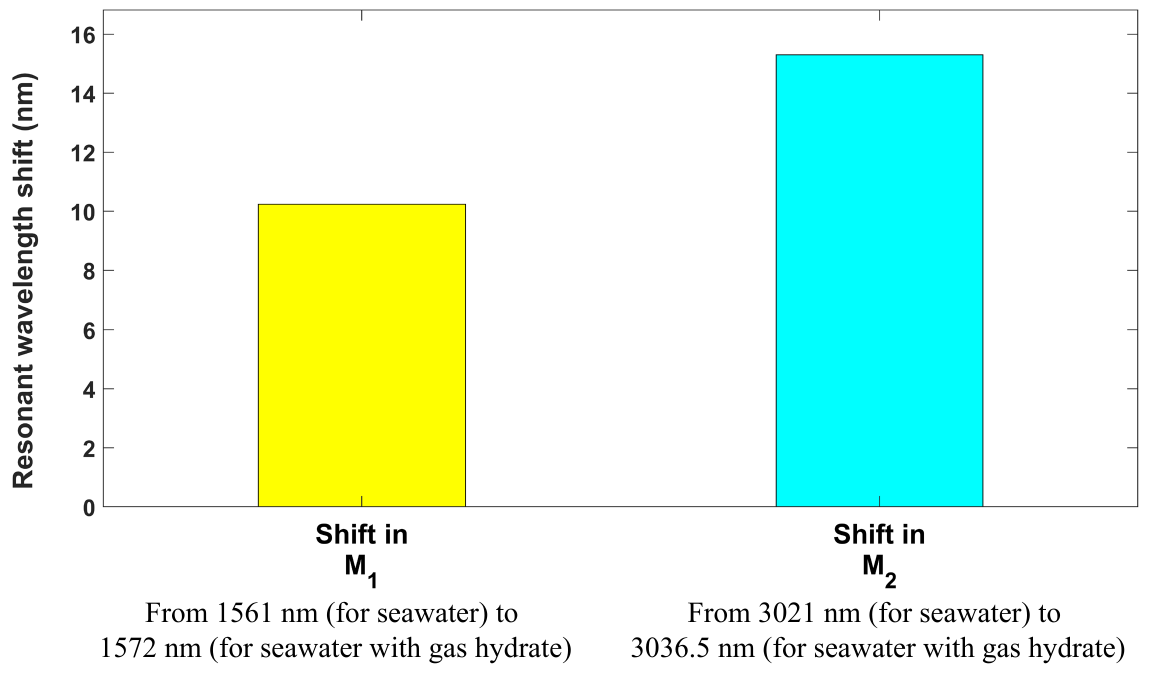}
    \captionsetup{labelformat=empty, labelsep=none}
    \caption{Figure S2: Shifts in the resonant wavelengths for both modes ($M_1$ and $M_2$) occur when transitioning from resonance in seawater to resonance in seawater containing gas hydrates.}
    \label{fig:Fig-s2}
\end{figure}

\noindent Simulation results show as in Figure S\ref{fig:Fig-s2} that the first resonant wavelength shifts from 1561 nm to 1572 nm and the second resonant wavelength shifts from 3021 nm to 3036.5 nm when the background refractive index changes from that of water to that of gas hydrate.


\subsection{Detection of different types of proteins in aqueous solutions}

\begin{figure}[H]
    \centering
    \includegraphics[width=1\textwidth]{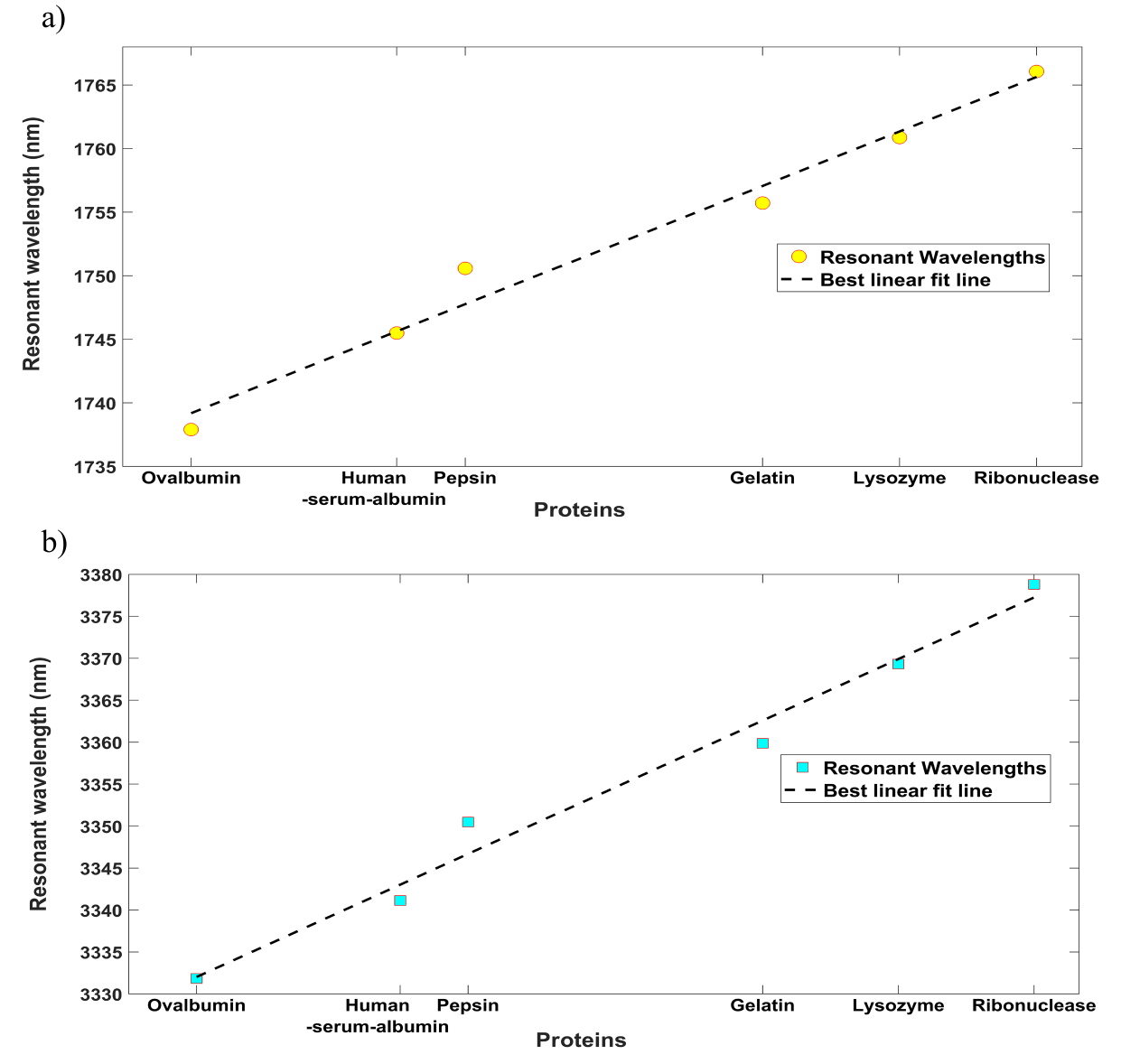}
    \captionsetup{labelformat=empty, labelsep=none}
    \caption{Figure S3: Resonant wavelengths for solutions (1\%-2\%) of ovalbumin, human serum albumin, pepsin, gelatin, lysozyme, and ribonuclease in water a) for $M_1$, b) for $M_2$}
    \label{fig:Fig-s3}
\end{figure}

We evaluated the effectiveness of our proposed structure in detecting different types of proteins in aqueous solutions. Different proteins in water, at specific concentrations, exhibit distinct refractive index values. Thus, the principle of refractive index sensing can be utilized to identify particular proteins in both water and alkaline solutions. Relevant data for protein solutions was collected from \cite{mcmeekin1964refractive}. We simulated the structure for 1\%-2\% aqueous solutions of five different proteins. Figure S\ref{fig:Fig-s3}.a), and b) show the corresponding resonant wavelengths for both modes of the absorption spectra from simulations considering solutions of ovalbumin, human serum albumin, pepsin, gelatin, lysozyme, and ribonuclease in water, respectively. Since the refractive index of protein solutions depends on the type of solvent, this method can also be applied to identify proteins dissolved in solvents other than water.

\subsection{Differentiation of amino acids from their isomers}
Differentiating amino acids from their isomers can be challenging due to the subtle structural differences that do not result in significant mass or charge variations\cite{lambeth2019differentiation}. Additionally, the molar refractions of amino acids and their isomers in a solution at a specific temperature are often nearly identical because they depend primarily on the molar mass of the solute. This similarity in molar refraction further complicates the differentiation of amino acids from their isomers in solution-based analyses\cite{arsule2020thermodynamic}. However, it has been found that isomers of amino acids in water exhibit different values of refractive index at a given temperature\cite{mcmeekin1964refractive}.

\begin{figure}[H]
    \centering
    \includegraphics[width=0.75\textwidth]{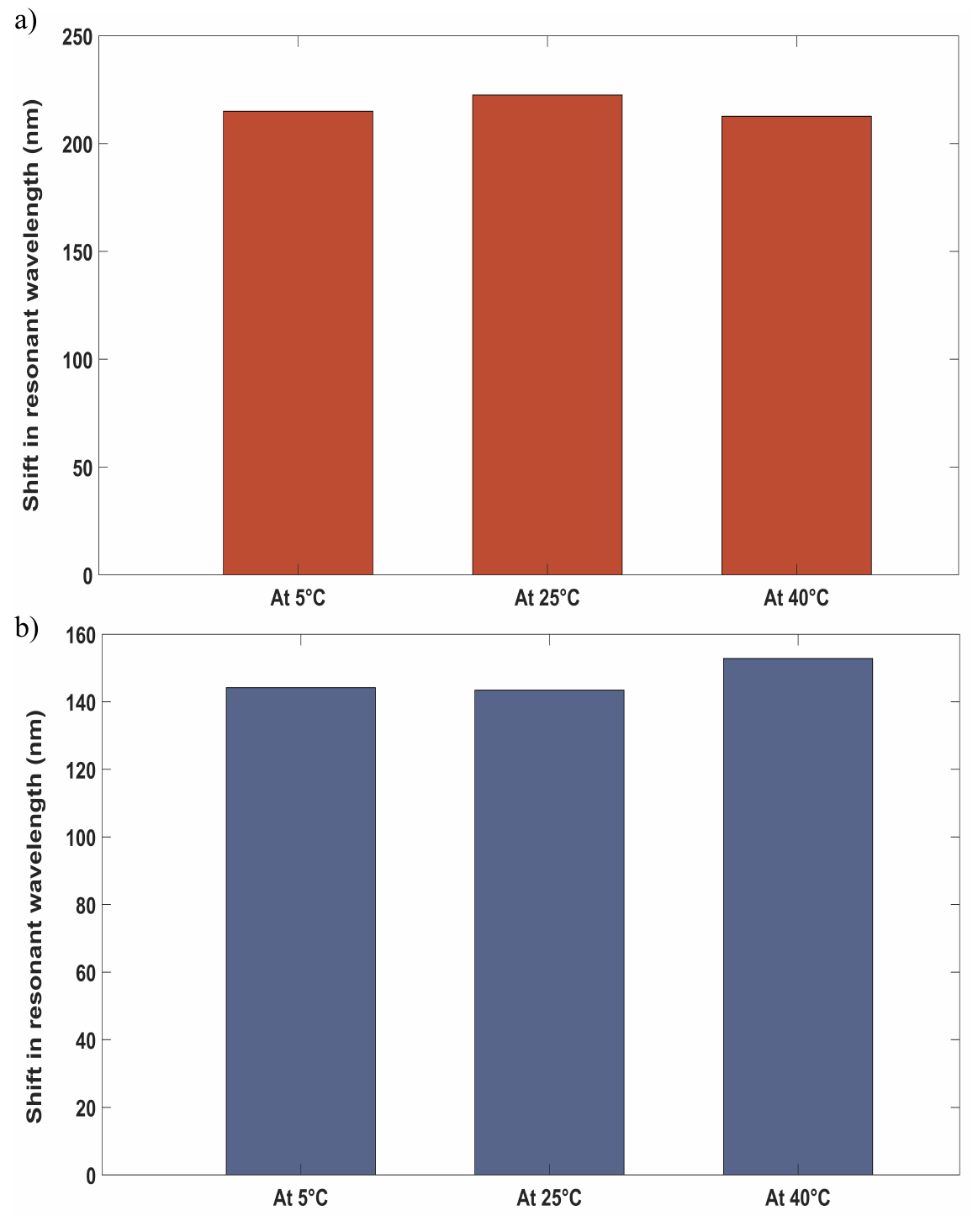}
    \captionsetup{labelformat=empty, labelsep=none}
    \caption{Figure S4: Shifts in the resonant wavelengths ($M_2$) for isomer pairs a) alanine-lactamide isomer pair, b) glycine-glycolamide isomer pair .}
    \label{fig:Fig-s4}
\end{figure}

\noindent We simulated our structure for two sets of isomers: glycine and its isomer glycolamide, and alanine and its isomer lactamide at three different temperatures. Results show the shifts in resonant wavelengths for these two amino acids and their isomers, as demonstrated in Figure S\ref{fig:Fig-s4}.a), and b). For this case, we only considered the shifts in the second peak of the absorption spectra.



\subsection{Determination of the concentration of various solutes in aqueous solutions}

\begin{figure}[H]
    \centering
    \includegraphics[width=0.82\textwidth]{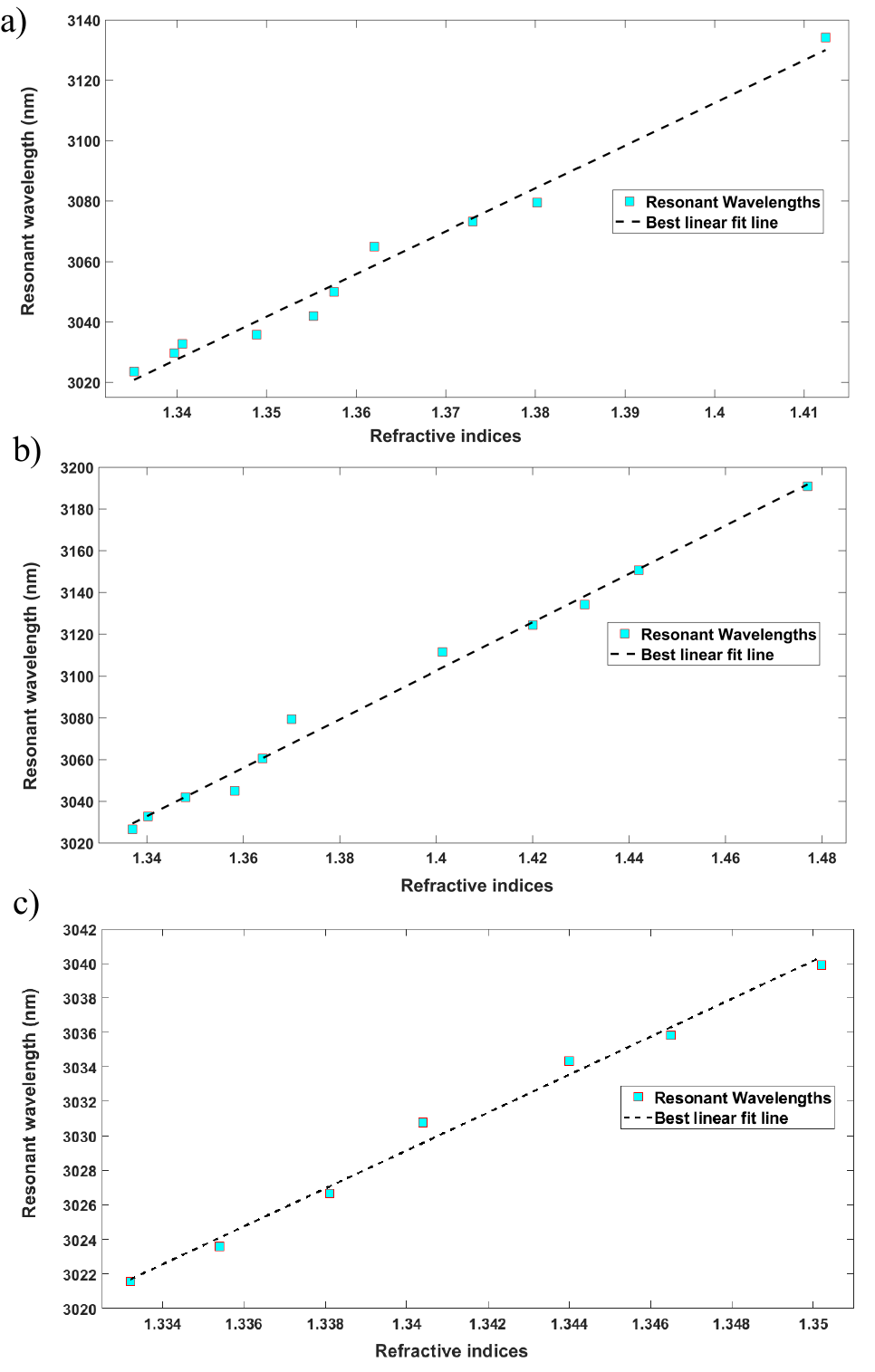}
    \captionsetup{labelformat=empty, labelsep=none}
    \caption{Figure S5: Resonance behavior of $M_2$ of the sensor for different concentrations of various solutes dissolved in water: a) for various inorganic salt solutions, b) for various sugar solutions, c) for ammonia solutions.}
    \label{fig:Fig-s5}
\end{figure}

We evaluated the compatibility and performance of our sensor in various aqueous solutions with differing concentrations. Since the refractive index is an indicator of solution concentration, we conducted simulations using various refractive index values, representing different solutions at varying concentrations. We are providing the findings for this application considering the second mode $M_2$ only. However, similar results were found considering the other mode, $M_1$, as can be verified by the response of the resonant wavelengths to the refractive indices provided in Figure 5.a).  First, we conducted simulations on aqueous solutions of various inorganic salts at different concentrations. We considered common salts, such as those containing Na\textsuperscript{+}, Ca\textsuperscript{+}, Fe\textsuperscript{3+}, NO\textsubscript{3}\textsuperscript{-}, SO\textsubscript{4}\textsuperscript{2-}, etc., which are commonly treated during water purification processes for drinking and industrial purposes\cite{guo2022nanofiltration,garg2022industrial}. Specifically, we used the corresponding refractive indices for sodium carbonate (Na\textsubscript{2}CO\textsubscript{3}) at 1\% and 3\% (1.3352, 1.3397), sodium nitrate (NaNO\textsubscript{3}) at 14\% and 40\% (1.3802, 1.3489), calcium chloride (CaCl\textsubscript{2}) at 10\% and 30\% (1.3575, 1.4124), ferric chloride (FeCl\textsubscript{3}) at 8\% and 14\% (1.3552, 1.3730), and sodium sulfate (Na\textsubscript{2}SO\textsubscript{4}) at 5\% and 20\% (1.3406, 1.3620)\cite{lide2004crc}. These results are illustrated in Figure S\ref{fig:Fig-s5}.a).

We also employed our structure to examine the resonance shifting trends in several common sugar-water solutions with varying concentrations. In this case, we used refractive indices of 1.348, 1.364, and 1.442 for D-glucose solutions at concentrations of 10\%, 20\%, and 60\%; for D-fructose (5\% and 24\%), the refractive indices were 1.3402 and 1.37; for lactose (0.5\% and 16\%), they were 1.337 and 1.3582; for maltose (40\% and 54\%), the values were 1.4013 and 1.4308; and for sucrose (50\% and 75\%), they were 1.42 and 1.477\cite{lide2004crc}. Figure S\ref{fig:Fig-s5}.b) provides an overview of the findings of this study.  
 
Lastly, we simulated different concentrations of ammonia dissolved in water and observed a similar resonance-shifting trend. Since an excess of dissolved ammonia poses a threat to aquatic life, and high levels of ammonia in drinking water can have detrimental effects on human health, accurately determining ammonia levels in water has become crucial in the agricultural, industrial, and public health sectors. Figure S\ref{fig:Fig-s5}.c) represents the results of our experiment to determine the concentration of dissolved ammonia, where we used refractive indices for ammonia in water solutions with concentrations ranging from 1\% to 30\%\cite{lide2004crc}.



\subsection{Detection of E. coli O157:H7 bacteria in food sample}

Escherichia coli O157:H7 has become a significant public health concern due to its role as an enteric pathogen. The infections it causes can present in various forms, from mild cases of diarrhea to severe and potentially life-threatening conditions like hemolytic uremic syndrome and hemorrhagic colitis. As this bacterium enters and invades the human intestine through contaminated food and drinks, the need for rapid detection methods to determine its presence in food items becomes crucial. We employed our sensor to detect E. coli O157:H7 in food and drink samples. To attach bacteria to the sensor surface, a layer of biotinylated polyclonal antibodies for E. coli O157
can be deployed, which is immobilized using neutravidin and phosphate-buffered saline (PBS) at pH 7.4. After antibody binding, PBS is rinsed to remove excess neutravidin and antibodies. Since the exact amount of binding is unpredictable, we assumed a bacteria-capturing layer thickness of 20 nm. In the simulation, this layer was modeled with a refractive index of 1.3435, which increases to 1.3465 when bacteria are captured\cite{waswa2007direct}. 

\begin{figure}[H]
    \centering
    \includegraphics[width=1\textwidth]{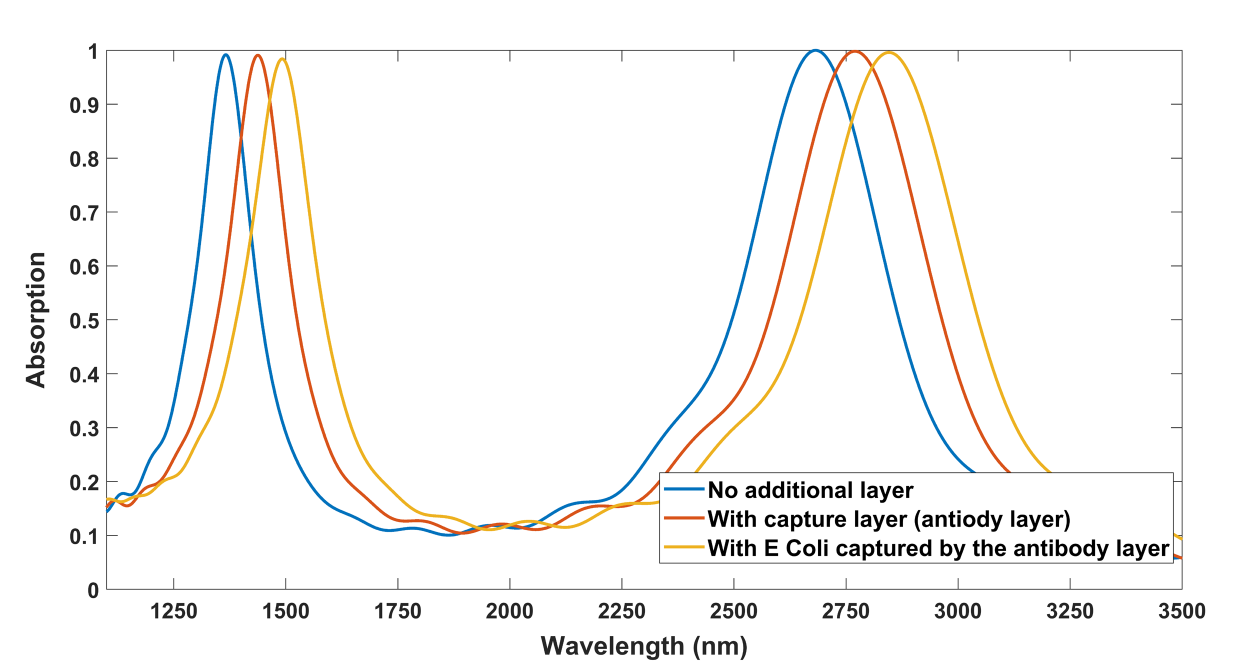}
    \captionsetup{labelformat=empty, labelsep=none}
    \caption{Figure S6: Absorption spectra for three different stages of the process of E. coli O157:H7 detection}
    \label{fig:Fig-s6}
\end{figure}

\noindent To imitate the captured E. coli O157:H7 ($10^5$ CFU/mL concentration), we again added a layer of 15 nm thickness on top of the antibody layer and changed the refractive index of both layers to 1.3465. Figure S\ref{fig:Fig-s6}. shows the absorption spectra for three scenarios: 1) with no additional layer added, 2) with only the antibody layer added, and 3) with bacteria attached to the antibody layer.



\subsection{Detection and determination of the amount of biomolecule attached to the polymer layer}

\begin{figure}[H]
    \centering
    \includegraphics[width=0.95\textwidth]{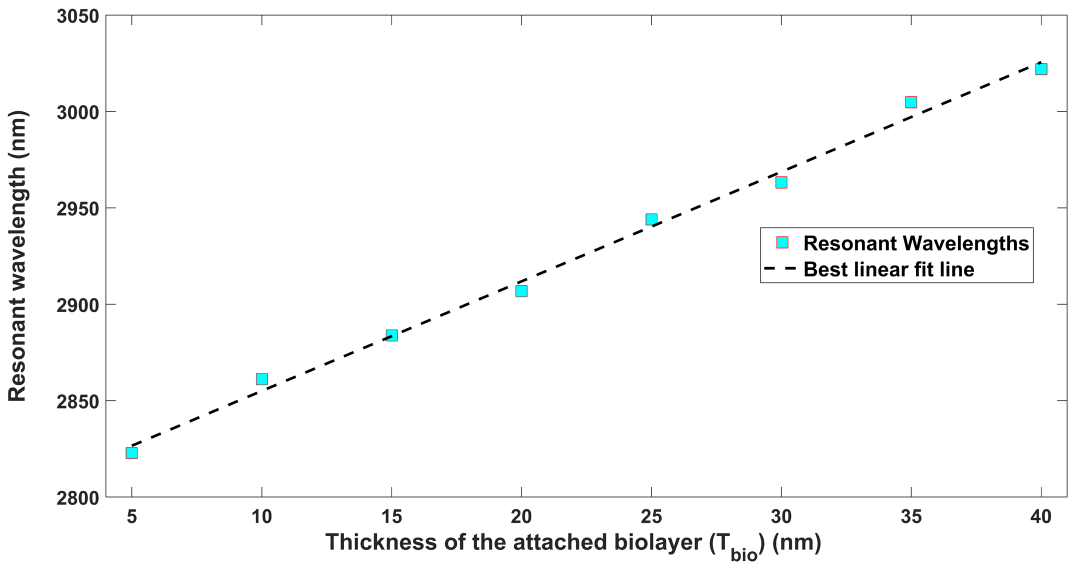}
    \captionsetup{labelformat=empty, labelsep=none}
    \caption{Figure S7: Resonance behavior of $M_2$ of the sensor in response to the changing thickness of the captured biolayer.}
    \label{fig:Fig-s7}
\end{figure}

We engaged our sensor for the detection of a particular antibody known as $\alpha$-streptavidin. Streptavidin and its homologs (collectively known as streptavidin) are extensively utilized in the biomolecular and bioengineering fields due to their highly selective and stable binding with biotin and because of their availability of diverse chemical and enzymatic biotinylation methods suited to various experimental setups\cite{dundas2013streptavidin}, while $\alpha$-streptavidin is an antibody commonly used to control, examine, and manipulate the amount of its antibody-antigen pair (streptavidin). In practical applications, streptavidin antibodies are anchored to the surface of biosensors via the familiar and highly selective antigen-antibody binding process\cite{jensen2005selective}. Different protocols, such as physical absorption or chemical coupling, can be used to attach a capturing layer of streptavidin antigen\cite{vesel2012immobilization}. The refractive index of streptavidin is typically chosen to be around 1.45, both in air and in a solvent, based on the analysis of streptavidin binding data from optical experiments\cite{weisser1999specific}. Although the refractive index and material dispersion of the biomolecular layer depend on the orientation of the biomolecules, we disregarded material dispersion in this case and assumed the refractive index of the attached biolayer of $\alpha$-streptavidin is 1.45 as well. The shifting trend of the resonant wavelength for $M_2$ with increasing thickness of the captured biolayer ($T_{bio}$ (assuming the thickness of the capture layer of antigen $T_{capture}$ = 25 nm) is shown in Figure S\ref{fig:Fig-s7}. 

\subsection{Detection of hemoglobin concentration in human blood}

We examined the sensing performance of our sensor in the detection of hemoglobin concentrations in human blood. Hemoglobin is a protein found in red blood cells, also called the erythrocytes, responsible for carrying oxygen to the body's organs and tissues while transporting carbon dioxide back to the lungs. Monitoring hemoglobin levels is crucial, as deviations from the normal range can indicate various health conditions. 

\begin{figure}[H]
    \centering
    \includegraphics[width=1\textwidth]{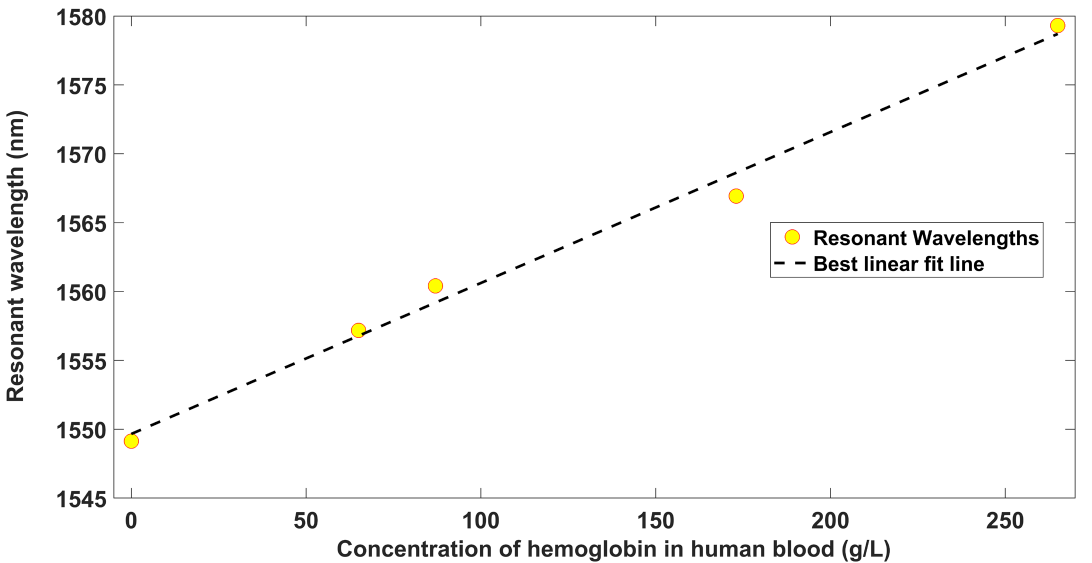}
    \captionsetup{labelformat=empty, labelsep=none}
    \caption{Figure S8: Resonance behavior ( for $M_1$) of the sensor in response to different densities of hemoglobin in the blood.}
    \label{fig:Fig-s8}
\end{figure}

\noindent To emulate the different concentrations of hemoglobin in the blood, we varied the refractive index of the surrounding media of the sensor and found a linear trend in shifting the resonance wavelength with the changing density of hemoglobin in the blood as illustrated in Figure S\ref{fig:Fig-s8}. Since a multiwavelength Abbe refractometer was used to measure the values of the refractive index of hemoglobin with various concentrations using light of 1550 nm, we considered the resonant wavelength of the first peak in our sensor’s absorption spectra\cite{lazareva2018measurement}. The linear fitting assumption we assumed here is justifiable, as the equation S\ref{equation:equation-s1} below can be used to calculate the refractive indices for different concentrations of hemoglobin.

\begin{equation}
    n = n_0 + \alpha C
    \label{equation:equation-s1}
\end{equation}

\noindent Here, the specific refraction increment of the refractive index, and the "effective" refractive index at zero concentration are denoted by $n_0$, and $\alpha$ respectively\cite{zhernovaya2011refractive}.



\subsection{Detection of DNA hybridization}

We also simulated our sensor to detect the completion of the process of DNA hybridization. During this process, single-stranded DNA (ssDNA) becomes double-stranded DNA (dsDNA), causing the refractive index of the DNA layer to shift from 1.456 to 1.53. This change in refractive index occurs due to variations in the density and polarizability of the molecule as ssDNA transforms into dsDNA. The increase in molecular mass associated with DNA hybridization contributes to this densification\cite{elhadj2004optical}. 

\begin{figure}[H]
    \centering
    \includegraphics[width=0.95\textwidth]{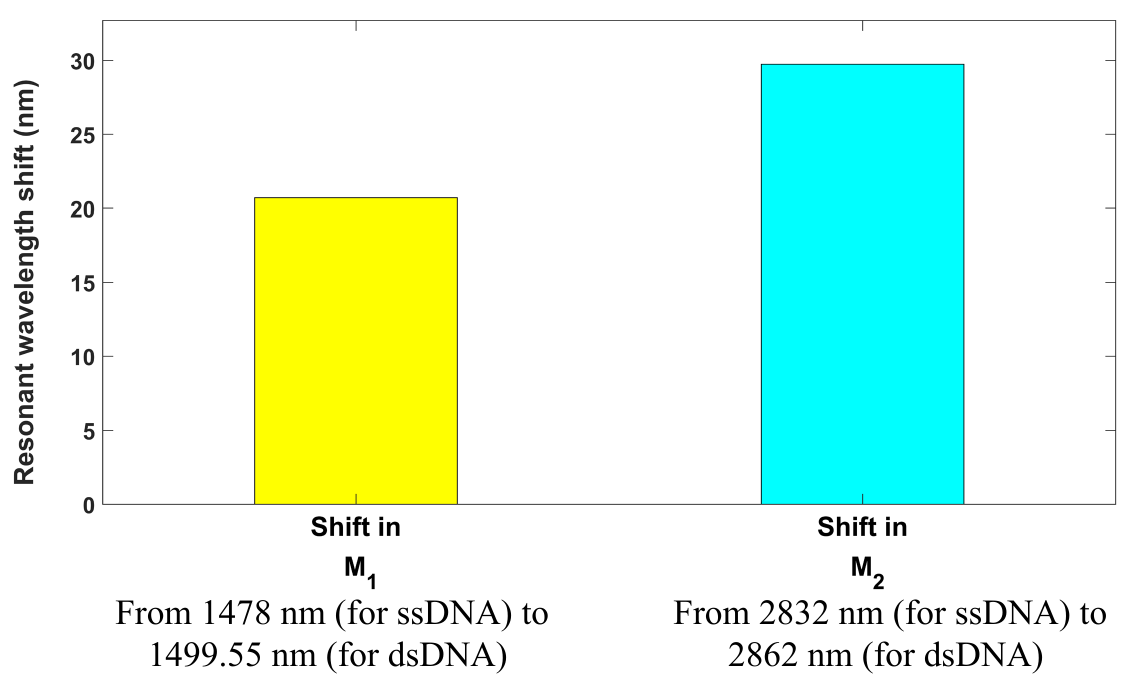}
    \captionsetup{labelformat=empty, labelsep=none}
    \caption{Figure S9: Shifts in the resonant wavelengths for both modes ($M_1$ and $M_2$) occur when transitioning from resonance before and after DNA hybridization.}
    \label{fig:Fig-s9}
\end{figure}

\noindent Initially, we applied an additional layer with a refractive index of 1.456 to represent the single-strand DNA (ssDNA) layer on the sensor, followed by another layer with a refractive index of 1.53 to simulate the double-strand DNA (dsDNA) layer after hybridization. Figure S\ref{fig:Fig-s9} demonstrates the simulation result that indicates a shift from 1478 nm to 1499.55 nm (a change of 21 nm) of the resonant wavelength of the first absorption peak and from 2832 nm to 2862 nm (a change of 30 nm) if we consider the second peak in the absorption spectrum. 



\subsection{Detection of different types of viruses
}

To assess the sensing compatibility further, we also utilized our sensor to detect various viruses individually. To simulate the attached viruses using virus immobilization methods, such as site-specific enzymes or hydrophobic polycations on the surface of the sensor, we applied an additional layer and later varied its refractive index to emulate different virus species\cite{meldal2016recent}. 

\begin{figure}[H]
    \centering
    \includegraphics[width=1\textwidth]{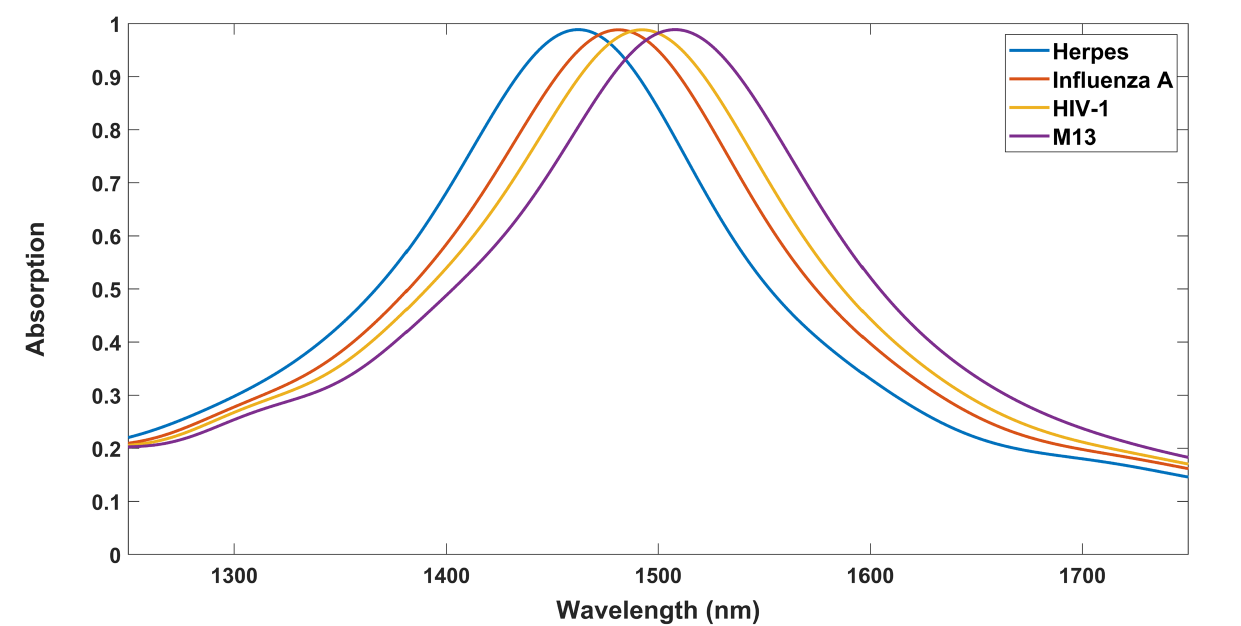}
    \captionsetup{labelformat=empty, labelsep=none}
    \caption{Figure S10: Absorption spectra of $M_1$ for different types of viruses attached to the surface of the sensor.}
    \label{fig:Fig-s10}
\end{figure}

\noindent Here, the viruses that we took into account are Herpes, Influenza A, HIV-1, and M13. The corresponding refractive indices to simulate these viruses were 1.41, 1.48, 1.5, and 1.57, respectively\cite{ymeti2007fast, wang2010label, block2012rapid, zhu2008opto}. Figure S\ref{fig:Fig-s10}. shows the absorption spectra considering the mode $M_1$ obtained from the simulations emulating these viruses attached to the sensor surface. 

\bibliographystyle{elsarticle-num}  
\bibliography{References}  